\documentclass[usegraphicx]{mn2e}

\usepackage{times}

\title[Red giant PL-relations in the LMC]{Red variables in the
OGLE-II database. I. Pulsations and period-luminosity relations
below the tip of the Red Giant Branch of the LMC}
\author[L. L. Kiss and T. R. Bedding]{L. L. Kiss\thanks{E-mail:
laszlo@physics.usyd.edu.au}\thanks{On leave from University of Szeged,
Hungary} and T. R. Bedding\\
\\
School of Physics, University of Sydney 2006, Australia}

\begin{document}

\date{Accepted ... Received ..; in original form ..}


\maketitle

\begin{abstract}

We present period-luminosity relations for more than 23,000 red giants in
the Large Magellanic Cloud observed by the OGLE-II microlensing project. 
The OGLE period values were combined with the 2MASS single-epoch $JHK_S$
photometric data.  
For the brighter stars we find agreement with previous
results (four different sequences corresponding to different modes of
pulsation in AGB stars). We also discovered two distinct and 
well-separated
sequences below the tip of the Red Giant Branch. They consist of almost
10,000 short-period (15 d $< P <$ 50 d), low-amplitude 
($A_{\rm I}<0.04$ mag) red variable stars, for which we propose that 
a significant fraction 
is likely to be on the Red Giant Branch, showing radial pulsations 
in the second and third overtone modes. The excitation mechanism could be 
either Mira-like pulsation or solar-like oscillations driven by convection. 

\end{abstract}


\section{Introduction}

Low- and intermediate mass stars (approximately 0.5 -- 5 M$_\odot$) enter late
evolutionary stages when they become red giants on the first giant branch
(RGB). Reaching the tip of the RGB (TRGB), rapid luminosity drop occurs after
the helium flash. The stars then expand again, climbing the Asymptotic Giant
Branch (AGB), where they  show strong pulsations (these are the Mira and
semiregular variables) associated with enhanced mass-loss. 

Recently, Ita et al. (2002) have reported the presence of many red variable
stars in the Large Magellanic Cloud around the TRGB. Based on their 
time-series $JHK$ photometry, they concluded that
besides faint AGB variables, a substantial fraction is likely to be on the RGB.
If confirmed, that would imply global pulsations do occur in a less evolved
stage than the AGB, which may have a strong impact on the pulsation-driven 
mass-loss history. 

Earlier, Welty (1985) performed a photographic search for RGB and AGB variable
stars in six globular
clusters, and found relatively high amplitude variability ($A_{\rm B}\leq 
0.2$ mag) only near the tip of the red giant branch. From 16 years 
of high-precision photoelectric photometry, Jorissen et al. (1997) found
a minimum-variability boundary for red giants, concluding that all stars 
of M spectral types are intrinsic variable stars. The observed radial velocity
jitter for some stars was interpreted as evidence for pulsationally induced
variability. This result on the absence of stable red giants was 
confirmed by Eyer \& Grenon (1997), who analysed the Hipparcos Epoch Photometry
database. From theoretical point of view, a thorough study was published 
by Dziembowski et al. (2001), who
presented a linear stability analysis of red giant models. One of their
conclusions was that if turbulent convection was included into the models, the
fundamental mode was strongly damped compared to higher overtones, i.e. the
more realistic models tended to have observable amplitudes for overtone modes
with orders $n\geq5$. That was in stark disagreement with the observations
of $\alpha$ UMa (Buzasi et al. 2000) and no definite
conclusion was drawn on the nature of possible excitation mechanisms. 
Possibilities included self-excitation of unstable modes (referred to
as Mira-like pulsations) and convection induced excitation of linearly
stable modes (the so-called solar-like oscillations). To make a distinction
between these, Dziembowski et al. (2001) suggested to re-analyse microlensing
data of the Magellanic Clouds to search for short-period and 
low-amplitude extensions of period-luminosity ridges (Wood et al. 1999) of red
variables. This paper presents the results of such a search.

Microlensing surveys have stimulated work in {\it ensemble asteroseismology},
which deals with global properties of pulsating stars (see Szabados \& Kurtz 2000
for reviews). AGB pulsators have been found to obey well-defined
period-luminosity (PL)  relations forming parallel ridges in the $\log P - K$
plane (Wood et al. 1999).  Wood (2000) discussed the nature of MACHO red
variables and concluded that the ridges  correspond to fundamental,
first, second and third overtone modes. His results  have been recently confirmed
by Cioni et al. (2001, 2003),  Noda et al. (2002) and Lebzelter et al. (2002). 

This Letter is the first in a series of papers dealing with an analysis 
of $I$-band observations obtained by the OGLE-II project (Zebrun et al. 2001).
Here we report on pulsating red giants in the Large Magellanic
Cloud and the existence of distinct PL relations 
for a large group of red variables below the TRGB. As shown below,
the most reasonable
explanation is the presence of global oscillations in a mixture of AGB and
RGB stars, probably in
the second and third overtone modes. A comparison between the Large and Small
Magellanic Clouds will be presented in a subsequent paper.
 
\section{Data analysis}

The second phase of the Optical Gravitational Lensing Experiment (OGLE-II,
Udalski et al. 1997)
spanned four years, from 1997 to 2000. About seven square degrees 
were observed repeatedly in the
Magellanic Clouds, yielding about 6 billion photometric measurements for about
20 million stars. The observations were reduced with a
modification of the Difference Image Analysis method (Alard \& Lupton 1998,
Wozniak 2000), which has been proved to be superior for detecting tiny 
photometric variations in crowded fields (the adopted 
definition of candidate variable stars was outlined by Wozniak 2000).
The $I$-band light curves typically contain 400 points per star.  
The error of individual magnitude measurements for the brightest stars 
($I<16$ mag) is a few mmags, so that fairly low-amplitude changes can be studied. 
More than 68,000 variable stars were listed in the OGLE-II catalog of variable
stars (Zebrun et al. 2001) and it forms the basis of this study.

There is one major advantage of using the OGLE-II database. It consists  of
$I$-band observations, so that red variables had relatively higher flux 
levels  in the original data than in ``bluer'' bands (such as, e.g. the MACHO
red band). Consequently, deeper limiting magnitudes and better photometric
precision is expected.  
On the other hand, there are two disadvantages. The first is the reduced
photometric amplitude of red giant pulsators in the $I$-band.  Because of
that, one
can expect difficulties in the low-amplitude regime. The other drawback is the
time-span of observations: a typical  OGLE-II light curve consists of 400
points obtained over 1200 days, which is less than half of the MACHO time-span.
As a result, period determination becomes uncertain for periods  longer than a
few hundred days.

We have found that, despite the smaller amplitudes in the $I$-band,
high-quality light curves enabled detection of periodicities of 
low-amplitude changes. 
For stars around $I=15-16$ mag, 1-percent variability could be safely 
analysed. For periods 
less than a year, period determination seems to be well-established.
Spurious periods occurred at integer multiples of a year, and they 
can be easily distinguished from real periods. 
For shorter periods, OGLE-II data are well-suited 
to infer accurate cycle lengths, and the possibility of studying
small amplitude stars outweighs the disadvantages.

\begin{figure}
\includegraphics[width=84mm]{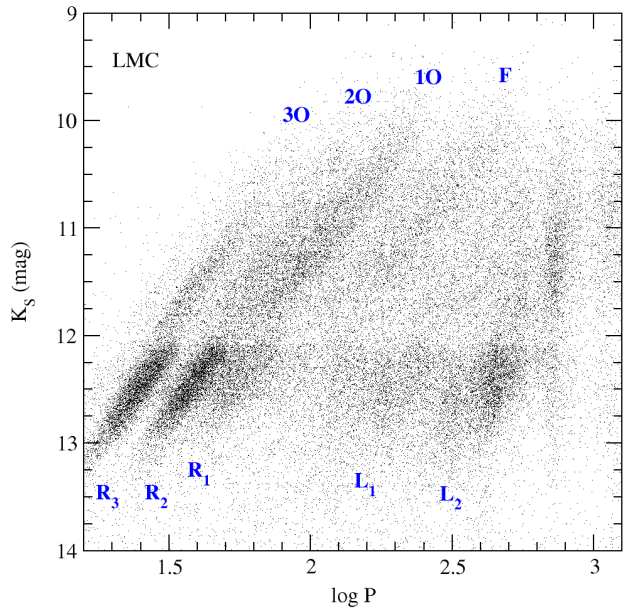}
\caption{PL relations for red giants in the LMC (62,591 periods for 
23,494 stars). This figure is available in colour in the on-line version
of the journal on {\it Synergy}.}
\label{plr}
\end{figure}

In order to reduce the effects of interstellar extinction and allow a
direct comparison with previous results, we focused on $\log P - M_{\rm K}$
relations. Our analysis consisted of the following steps:

\begin{enumerate}

\item We made a cross-correlation of the 52,937 OGLE-II variable stars
in the LMC with the 2MASS All-Sky Point Source Catalog\footnote{\tt 
http://irsa.ipac.caltech.edu}; with 
a search radius of 1$^{\prime\prime}$ we found 32,062 stars with $JHK_S$
magnitudes.

\item Next, we excluded the negligible number (7) of duplicates 
(i.e. when two stars were found within 1$^{\prime\prime}$ of 
the OGLE coordinates).

\item Red giants were selected by the $J-K_S$ colour; as a safe threshold we 
chose 0.9 mag (according to standard lists of Hawarden et al. 2001, 
that corresponds to early M spectral type). The final sample
consisted of 23,494 stars.

\item We performed a period search on OGLE light curves by an iterative
Fourier analysis. First, we 
calculated the Discrete Fourier 
Transform (the examined frequency range was between 0 d$^{-1}$ and 0.066 
d$^{-1}$ with a frequency step of $6\times10^{-7}$ d$^{-1}$). After finding
the highest peak in the spectrum, we subtracted
a sine wave from the data with that amplitude and the optimal phase. Then
the whole procedure was repeated until a four-component fit was 
reached. We kept only those frequencies larger than 
$8\times10^{-4}$ d$^{-1}$ ($\sim$1/T$_{\rm obs}$) and with semi-amplitudes 
larger than 5 mmag. 

\item For each star, the resulting database contains the OGLE identifier
(made of the J2000.0 coordinates), periods, semi-amplitudes, phases, mean $I$ 
magnitude and 2MASS $JHK_S$ single-epoch magnitudes. 

\end{enumerate} 

There are some necessary simplifications in this procedure enforced by the
large number of stars. However, it is exactly this factor that enables
good statistics. In overlapping regions between previous works and 
our study (brighter stars with larger amplitudes), the agreement is excellent.
For instance, Cioni et al. (2001, 2003) made very careful, object-to-object
control of light curves before studying statistics, and their results 
are very well reproduced by our data. The tight constraint on coordinate 
made very likely that infrared counterparts have been unambiguously identified.
Random checks of period determination showed that below 200 days, periods are
likely to be accurate to a few percent. Five percent period error 
($\delta \log P$=0.02) is smaller than the intrinsic period jitter well-known 
in local semiregular variables (Kiss et al. 1999). For longer periods, there 
are unavoidable groups of
spurious periods at integer multiples of a year, which are easily recognizable
as vertical strips in the period-luminosity plane. We consider our data 
to be reliable up to about 500--600 days, for which at least two cycles 
were covered by the observations. 
  
\section{Discussion}

\begin{figure}
\includegraphics[width=84mm]{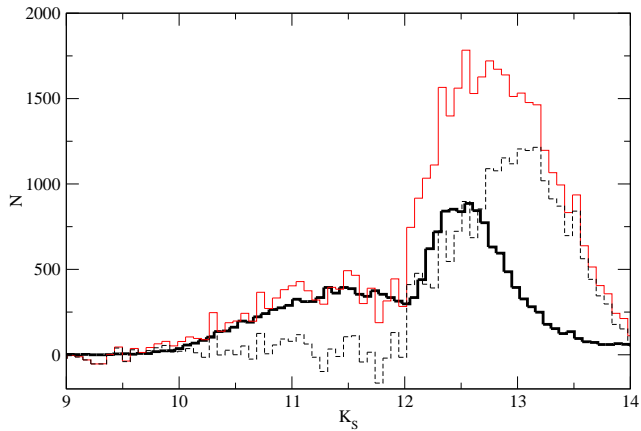}
\caption{Stellar magnitude distributions for OGLE-II variables, 
(thick solid line), LMC foreground-subtracted sample from Cioni et al.
(2000) (thin 
red solid line) and the difference (dashed line). This figure is 
available in colour in the on-line version
of the journal on {\it Synergy}.}
\label{lf}
\end{figure}

We present the resulting PL relations in Fig.\ \ref{plr}. The
outstanding features of the diagram: {\it (i)} there is a sudden drop 
of stellar density at $K_S=12.05$, located exactly at the TRGB 
(see Cioni et al. 2000 for details); {\it (ii)} above the TRGB, there are 
four sequences (two partly overlapping)
identified by Wood (2000) as stars pulsating in fundamental (F), first (1O),
second (2O) and third (3O) overtone modes (in his notation, these were sequences
C, B and A); {\it (iii)} below the TRGB, there are
two distinct and well-separated sequences (R$_2$ and R$_3$), 
slightly shifted relative to 2O and 3O; {\it (iv)} there is also a continuation
of 1O below the TRGB (R$_1$), less populated by an order of magnitude 
than R$_2$ and R$_3$; 
{\it (v)} there are two  long-period sequences
(extending well below the TRGB), one of which has been explained by Wood (2000) 
as binary variables (L$_1=$E in Wood 2000) and the other being long secondary 
periods of ambiguous origin (L$_2=$D, see also Olivier \& Wood 2003). Similar
ridges and distributions have also been found
for the SMC (Kiss \& Bedding, in prep.), which means we did not find 
strong dependence on metallicity, at least over the range covered by 
the LMC and SMC. In the following, we discuss some general properties 
with special emphasis on variables below the TRGB. In order to reveal some
hidden features in Fig.\ \ref{plr}, we have examined 
the luminosity function (LF), colour-magnitude diagram (CMD) and
the dependence on amplitude (including colour information).

\begin{figure}
\includegraphics[width=84mm]{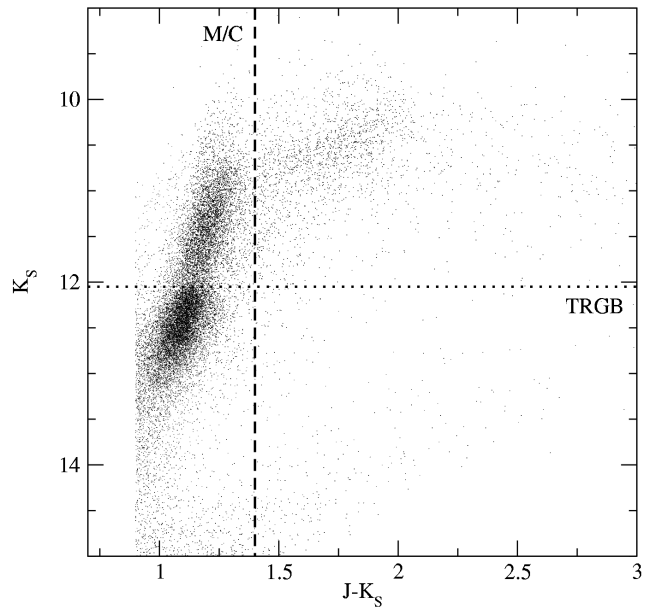}
\caption{The colour-magnitude diagram. Dashed line shows 
the boundary between oxygen-rich and carbon stars, while dotted line represents
the TRGB.}
\label{cmd}
\end{figure}

\begin{figure*}
\includegraphics[width=160mm]{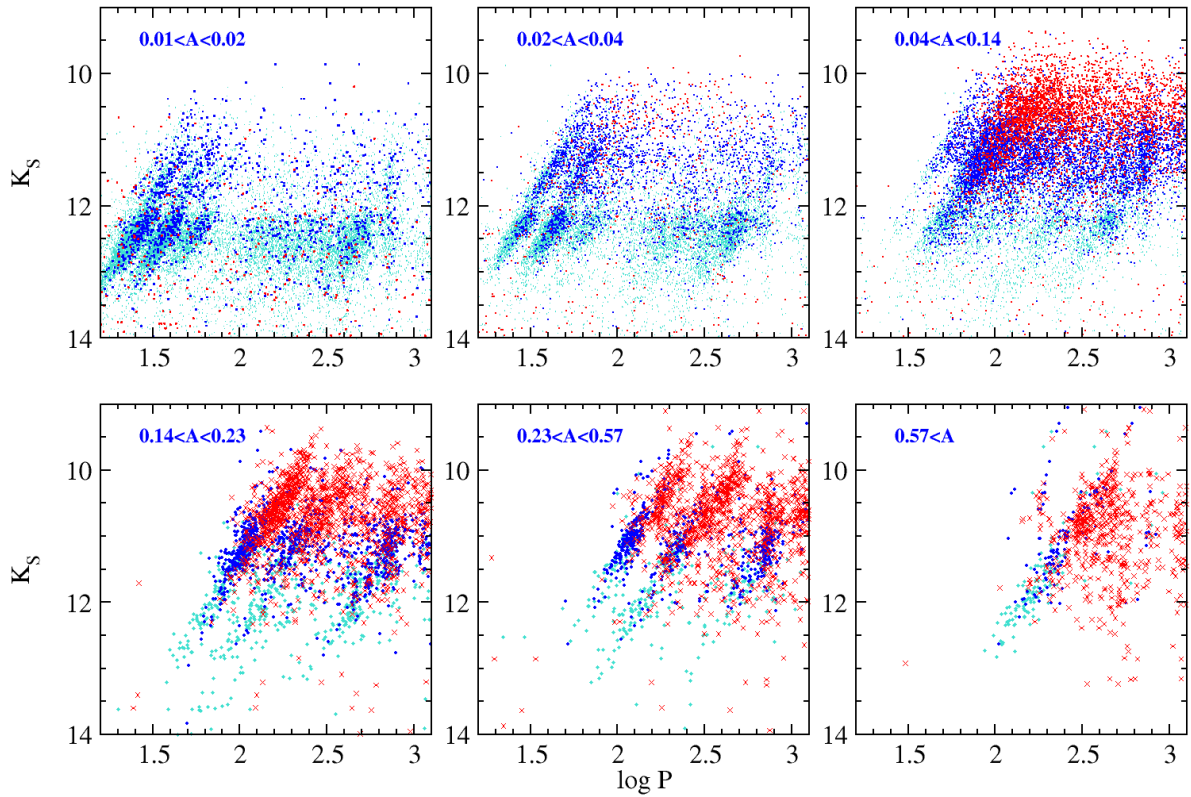}
\caption{PL relations in the LMC as function of the full amplitude of modes.
Three $J-K_S$ colour ranges were selected to plot in different colours.
Turquoise (light gray): $0.9<J-K_S\leq1.2$; blue (black): $1.2<J-K_S\leq1.4$; 
red (dark gray): $J-K_S>1.4$. This figure is available in colour in the on-line
version of the journal on {\it Synergy}. Symbols in
the lower panels are drawn larger for improved clarity.}
\label{lmcplra}
\end{figure*}

The LF (thick solid line in Fig.\ \ref{lf}) is clearly bimodal, with two
approximately Gaussian components. A least-squares fit gives 
K$_1$=11.33$\pm$0.05 mag, K$_2$=12.54$\pm$0.01 mag, $\sigma_1$=1.81$\pm$0.14 mag
and $\sigma_2$=0.82$\pm$0.03 mag for the maxima and full-widths at half maxima,
respectively. The boundary between them is at $K_S\approx12.0$ mag, which is in
excellent agreement with the TRGB value determined by Cioni et al. (2000). For
this, two possible explanations exist. Alves et al. (1998) and Wood (2000) 
suggested that all variables below the TRGB are thermally pulsing AGB stars and
their brighter cut-off just coincides with the TRGB. This has been  questionned
by Ita et al. (2002), who argued that a significant fraction  is likely to
consists of first-ascent red giants. To show what level of coincidence is
required for the former interpretation, we compare our  LF with the more complete
sample of Cioni et al. (2000) (thin line  in Fig.\ \ref{lf}). Their data are
foreground-subtracted DENIS magnitude distribution, which means we do not have
information on the spatial distribution of stars in the sample, and so we did not
attempt to correct for the different survey areas. Two features are worth
noticing: {\it (i)} the bright AGB components of the two LFs are essentially the
same, so that OGLE-II has detected practically all AGB stars above the TRGB; {\it
(ii)} the rise at $K_{\rm S}=12$ occurs in exactly the same place in both LFs (to
within 0.07 mag). The latter point is strong evidence that the OGLE variables
below the TRGB contain substantial numbers of RGB stars. The colour distributions
(see below) will give further empirical support to  this argument.

The $K_S$ vs. $J-K_S$ CMD (Fig.\ \ref{cmd}) looks very similar to the 
one presented in Cioni \& Habing (2003), which was based on a colour-selected 
sample. 
The overall agreement shows there is no significant difference between 
variability-selected and colour-selected samples. We show the stellar density
drop estimated from Figs.\ \ref{lf} (a more sophisticated 
determination is beyond the scope of the present paper) and the 
conventional boundary between oxygen-rich and carbon-rich stars 
($J-K_S=1.4$ mag). Stars below the TRGB are well concentrated in 
the range $J-K_S=0.95-1.3$ mag, so that
there is no excess of some redder and fainter population of stars, which
might be associated with, for instance, obscured AGB stars 
(van Loon et al. 1998) or low-mass carbon stars below the TRGB (Lattanzio 1989)
recognizable from their extreme red colours. 
 
In Fig.\ \ref{lmcplra}, we plotted six different slices 
of the (period, amplitude, $K_S$ magnitude) data cube, where  
the amplitudes were calculated as twice of the Fourier semi-amplitudes.
To reveal more information, we also indicate three different $J-K_S$ colour ranges
($J-K_S=0.9-1.2$, $1.2-1.4$ and $>1.4$ mag). For normal M-type stars (with no
dust), $\delta(J-K)=0.1$ mag corresponds to $\delta T_{\rm eff}\sim200$ K  
(Bessell et al. 1998), thus the chosen ranges enable a rough classification 
of stars as `hot', `warm' and `cool' red giants (actually, the latter
group is likely to consist of carbon stars). The exact temperature scale is
not important for our purposes, because we are only interested in their 
distribution. 

There are many interesting features in Fig.\ \ref{lmcplra}. 
Firstly, there is a sudden disappearance of stars below the TRGB with
increasing amplitude. For amplitudes larger than 0.04 mag, 
very few short-period ($P<50$ d) stars remain, while for $A>0.14$, almost every
star fainter than the TRGB disappears. The lower three panels show very clearly
the fundamental and first overtone sequences and hints of the 
long secondary period sequence. The upper panels are dominated by the higher
overtone pulsators, so that there is a good correlation between the amplitude
and mode of pulsation. Secondly, the colour distributions show very well
the temperature differences within each mode of pulsation. The lower three
panels are in good agreement with the expectations: for a given mode and
assuming a narrow range of masses, the
PL relation is equivalent to a density-luminosity relation,
which implies a monotonic temperature variation along any particular sequence. 
The upper-right panel reveals that stars in the overlapping 1O and 2O ridges 
are easily distinguishable by their colours and the transition between
them is quite sharp (1O contains very few blue dots).
Thirdly, ridges R$_1$, R$_2$ and R$_3$ differ from the other ridges
in that each contains both `hot' (turquoise) and `warm' (blue) red giants 
along their full extent, with a tendency for `warm' stars to have 
slightly longer periods (i.e. the blue dots are concentrated on the
right side of the ridges).

We interpret this last point as evidence for RGB stars mixing with thermally
pulsing AGB variables. 
At the TRGB, RGB and AGB stars have very similar luminosities,
with a tendency for RGB stars to have slightly lower temperatures. 
Recent evolutionary models of the Magellanic Clouds 
(Castellani et al. 2003) show that for stars with 1--2 M$_\odot$, the
temperature difference at constant luminosity is  
$\delta \log T_{\rm eff}\approx0.01$. This leads to  
$\delta \log R\approx-0.02$, or $R_{\rm AGB}/R_{\rm RGB}\approx0.96$. For a
given mode of pulsation, the period-density relation ($P\sqrt{M/R^3}=Q$) results 
in $\delta \log P=1.5\times\delta \log R\approx-0.03$, in good 
agreement with the period shift $\sim-0.05$ within R$_2$ and R$_3$ in
the upper left panel of Fig.\ \ref{lmcplra}. On the other hand, 
the fact that R$_2$ and R$_3$ are apparent continuations of 2O and 3O
ridges, suggests they are second and third overtone pulsators, a mixture of
TPAGB and RGB stars. We also note that even TPAGB stars show a slight 
shift relative to 2O and 3O, which is in agreement with their higher
temperatures in lower luminosity regions (Vassiliadis \& Wood 1993).
The apparent displacement of ``hot'' and ``warm'' red giants is 
therefore consistent with the assumption of the same mode and 
we conclude RGB stars do pulsate mostly in the second and third overtone 
modes (some penetration is also likely in R$_1$, the faint continuation 
of the first overtone ridge).

Finally, we also note in the lower right panel an interesting group of Mira 
stars located below the PL sequence. We identify them (Kiss, in
prep.) with dust-enshrouded Mira stars (Wood 1998).

\section{Conclusions and future work}

In this paper we presented the first results of a complex analysis of
OGLE-II red variable stars in the LMC. Our most important result is 
the discovery 
of separate PL relations of a huge number of variable stars 
below the TRGB. We found 9,617 stars with $K_S>12.05$ mag
having at least one period shorter then 50 days (that is 41\% of the full sample).
Alves et al. (1998), Wood et al. (1999) and Wood (2000) also detected 
such stars in the MACHO data (though 
their samples were more than an order of magnitude smaller) and 
they favoured thermally pulsing AGB explanation, instead of 
placing stars on the RGB. However, our sample is the largest available to date
and the extreme number of variables below the TRGB can be hardly explained
solely by accumulated TPAGB (or early AGB) stars at the TRGB. The revealed
colour dependence, accompanied with the slight period shift, can be consistently
explained by the mixture of AGB and RGB pulsators. 

The existence of separate
PL relations below the TRGB leads to the conclusion that RGB pulsations 
have remarkable astrophysical potential. On one hand, they extend 
applicability of asteroseismological considerations to a so-far neglected 
class of stars. On the other hand, their PL relations can serve as a powerful
test of globular cluster (GC) distance scale. The brighter ends of the RGB
luminosity functions (Zoccali \&
Piotto 2000) suggest that the number of upper-RGB stars in typical GCs
can be from tens to hundreds, so that
reliable statistics might be obtained within a reasonable time-span. This is
provided by the  short periods (15 to 50 days) and relatively large
photometric  amplitudes (0.01 to 0.04 mag in $I$-band). We will address 
the problem of separating RGB stars in R$_2$ and R$_3$ and 
derivation of useful relations for future GC observations in a forthcoming
publication. 

An interesting question is, where are the local counterparts of the RGB
pulsators? One of the closest red giant stars is $\gamma$ Crucis  (M3III, d=27
pc, ESA 1997), for which radial velocity data by Cummings (1999) suggested a
pulsational time scale of 13--16 days ($\log P\approx1.17$). Its 2MASS
$K_S=-3.258$ mag translates to M$_{\rm K_S}=-5.41$ mag, or $K_S$(LMC)=13.09 mag
(adopting  $\mu$(LMC)=18.50). This position is in excellent agreement with R$_3$.
Although this is only a single  example, is it reasonable to predict similar
behaviour for other local  RGB stars, too. 

With the presented properties, RGB pulsators bridge the gap between the
AGB pulsators and low-amplitude K giant pulsators discovered in 47
Tucanae by Edmonds \& Gilliland (1996). The emerging view of red giant (RGB and
AGB) pulsations suggests that all stars do pulsate similarly, in the sense of
obeying similar PL-relations. Only the mode distribution is changing over the
varying physical parameters (and possibly evolutionary status). The majority of
stars  oscillates in high overtones, which are heavily
perturbed by, e.g., the convection, leading to the ``semiregular'' behaviour.
However, they remain close relatives  of the more regular Mira stars, with
different modes of pulsation. In a future work 
we will tackle the problem of photometric
mode identification exploiting the ensemble properties of red giant
light curves to gain a better understanding of mode distributions in 
local variable stars.

The OGLE-II database provides an excellent opportunity to
study general properties of red giant pulsators, being the most extensive 
sample presently available. There is a number of questions which may strongly
benefit from utilizing the statistical power of OGLE-II data. They include, for
instance, a very accurate relative distance modulus of the LMC and SMC, a
robust comparison of pulsational properties in the Magellanic Clouds and the
Bulge (Wozniak et al. 2002) and finding evolutionary effects on pulsation.
By analysing OGLE-II data, a 
real advance is expected in theoretical understanding of red giant stars,
being a major issue of the contemporary stellar astrophysics.
 
\section*{Acknowledgments} 

{\footnotesize This work has been supported by the FKFP Grant
0010/2001, OTKA Grant \#F043203 and the Australian Research Council. 
Thanks are due to an anonymous referee, whose criticism and suggestions
led to significant improvement of the paper. We also thank Dr. M.-R.L. Cioni
for providing DENIS luminosity function plotted in Fig.\ \ref{lf}.
This
research has made use of the NASA/IPAC Infrared Science Archive, which is
operated by the Jet Propulsion Laboratory, California Institute of Technology,
under contract with the National Aeronautics and Space Administration. 
The NASA ADS Abstract Service was used to access data and references.}

\end{document}